\documentclass[twocolumn,english,superscriptaddress,prl,amsmath,amssymb,floatfix,longbibliography,nofootinbib]{revtex4-2}
\usepackage[T1]{fontenc}
\usepackage[utf8]{inputenc}
\usepackage{color}
\usepackage{babel}
\usepackage{amsmath}
\usepackage{graphicx}
\usepackage{esint}
\usepackage[unicode=true,pdfusetitle,
 bookmarks=true,bookmarksnumbered=false,bookmarksopen=false,
 breaklinks=false,pdfborder={0 0 0},pdfborderstyle={},backref=false,colorlinks=true]
 {hyperref}

\makeatletter

\usepackage{dcolumn}
\usepackage{bm}
\usepackage{color}
\usepackage{soul}
\usepackage{babel}
\usepackage{amsfonts}
\usepackage{slashed}
\usepackage{enumerate}

\usepackage{babel}

\makeatother

\begin{document}
\definecolor{orange}{rgb}{1.0, 0.5, 0.0}
\def\yuvalgood#1{\textcolor{blue}{#1}}%
\def\yuvalbad#1{\textcolor{orange}{#1}}%
\def\kyrylo#1{\textcolor{cyan}{#1}}%
\global\long\def\sgn{\mathrm{sgn}}%
\global\long\def\ket#1{\left|#1\right\rangle }%
\global\long\def\bra#1{\left\langle #1\right|}%
\global\long\def\sp#1#2{\langle#1|#2\rangle}%
\global\long\def\abs#1{\left|#1\right|}%
\global\long\def\avg#1{\langle#1\rangle}%

\title{Topological transitions of the generalized Pancharatnam-Berry phase}
\author{Manuel F. Ferrer-Garcia}
\affiliation{Nexus for Quantum Technologies, University of Ottawa, K1N 5N6, ON, Ottawa, Canada}
\author{Kyrylo Snizhko}
\affiliation{Department of Condensed Matter Physics, Weizmann Institute of Science,
Rehovot, 76100 Israel}
\affiliation{Institute for Quantum Materials and Technologies, Karlsruhe Institute
of Technology, 76021 Karlsruhe, Germany}
\affiliation{Univ. Grenoble Alpes, CEA, Grenoble INP, IRIG, PHELIQS, 38000 Grenoble,
France}
\author{Alessio D'Errico}
\email{aderrico@uottawa.ca}
\affiliation{Nexus for Quantum Technologies, University of Ottawa, K1N 5N6, ON, Ottawa, Canada}

\author{Alessandro Romito}
\affiliation{Department of Physics, Lancaster University, Lancaster LA1 4YB, United
Kingdom}
\author{Yuval Gefen}
\affiliation{Department of Condensed Matter Physics, Weizmann Institute of Science,
Rehovot, 76100 Israel}
\author{Ebrahim Karimi}
\affiliation{Nexus for Quantum Technologies, University of Ottawa, K1N 5N6, ON, Ottawa, Canada}
\email{ekarimi@uottawa.ca}

\date{\today}
\begin{abstract}
Distinct from the dynamical phase, in a cyclic evolution, a system’s state may acquire an additional component, a.k.a. geometric phase. The latter is a manifestation of a closed path in state space. Geometric phases underlie various physical phenomena, notably the emergence of topological invariants of many-body states. Recently it has been demonstrated that geometric phases can be induced by a sequence of generalized measurements implemented on a single qubit. Furthermore, it has been predicted that such geometric phases may exhibit a topological transition as function of the measurement strength. Here, we demonstrate and study this transition experimentally employing an optical platform. We show the robustness to certain generalizations of the original protocol, as well as to certain types of imperfections. Our protocol can be interpreted in terms of environment-induced geometric phases.
\end{abstract}
\maketitle

\section{Introduction}


When a quantum state undergoes a cyclic evolution, the phase acquired is given by the well-known dynamical component plus an additional contribution, associated with the geometrical features of the path followed by the state. This additional contribution is known as the geometric phase. The general framework for the emergence of a geometric phase has been pointed out first by Berry \cite{Berry1984} in the context of adiabatic quantum evolution. A specific realization of this phase had been earlier considered by Pancharatnam \cite{Pancharatnam1956} in his study of generalized interference theory. Pancharantnam's theory shows how geometric phases can be acquired in a non-adiabatic cyclic evolution, noting that these are given by the area enclosed by the respective trajectory of the system in the state space. The Pancharantnam phase can be observed following a sequence of running projective measurements, each of a different observable, where the last measurement projects on the initial state \cite{Berry1996,Chruscinski2004}. Geometric phases have found applications in several fields of physics~\cite{Cohen2019}, in particular, in optics~\cite{bomzon2002space,bliokh2019geometric,rubano2019q,jisha2021geometric} and condensed matter physics~\cite{LH:58,zak1989berry,resta2000manifestations,nayak2008non,fradkin2013field}. Importantly, the Berry phase is a key theme for understanding topological phases of matter \cite{fradkin2013field}. For instance, the Berry phase plays the role of a topological invariant in one-dimensional chiral symmetric systems~\cite{resta1994macroscopic,asboth2016short} and serve as the fundamental building block in the definition of other topological invariants, such as Chern numbers \cite{hasan2010colloquium}. Going beyond Hamiltonian dynamics, the emergence of geometric phases has been predicted and observed in the context of non-Hermitian evolution~\cite{garrison1988complex,dattoli1990geometrical,el2018non}; such phases were further shown to emerge following a sequence of weak measurements~\cite{Cho2019,Gebhart2020}. Further pursuing the latter theme, a major theoretical development has revealed that dynamics comprising multiple measurements may assign topological features to geometric phases. In particular, the limits of weak and strong measurement are topologically distinct~\cite{Gebhart2020,Snizhko2021,Snizhko2021a}. This prediction has recently been confirmed employing a superconducting qubit platform~\cite{Wang2021}. That study has implemented postselection on each individual measurement. This aligns with the original theoretical proposal~\cite{Gebhart2020,Snizhko2021,Snizhko2021a}, yet it leaves the question open: \textit{to what extent is the predicted topological transition a feature of the specific laid-down protocol?}

The experiment reported here not only employs a platform different from that of Ref.~\citep{Wang2021} (namely, an optical platform),  but also introduces a protocol which is conceptually different: rather than exercising postselection on each individual detector's readout, here we implement postselection on a joint readout of {\it all} measurements of the run. We  find that a topological phase transition takes place also under such generalized conditions, with distinct values of the topological number characterizing the respective limits of projective and infinitely weak measurements. 
Our experimental procedure consists of a sequence of measurements, each implemented by a set of optical elements. The key optical element is a polarization-sensitive beam displacer, which is used to execute a weak measurement on the polarization state of a laser beam. Employing additional elements (quarter wave plates, compensating wave plates) serves to tune the measurement to a specific observable. The strength of the measurement is determined by the ratio of the beam width and the difference of transverse displacements of orthogonal polarizations. Importantly, the detector’s readout is, in fact, the polarisation degree of freedom of the photon, which, in turn, could be viewed as the system, while the transverse position can be viewed as the environment. Our protocol could then be interpreted as an environment inducing a geometric phase, highlighting the dual nature of detector/environment. 
Finally, we investigate the robustness of the observed topological properties with respect to setup imperfections.

\section{Results}

\begin{figure}
\begin{centering}
\includegraphics[width=1\columnwidth]{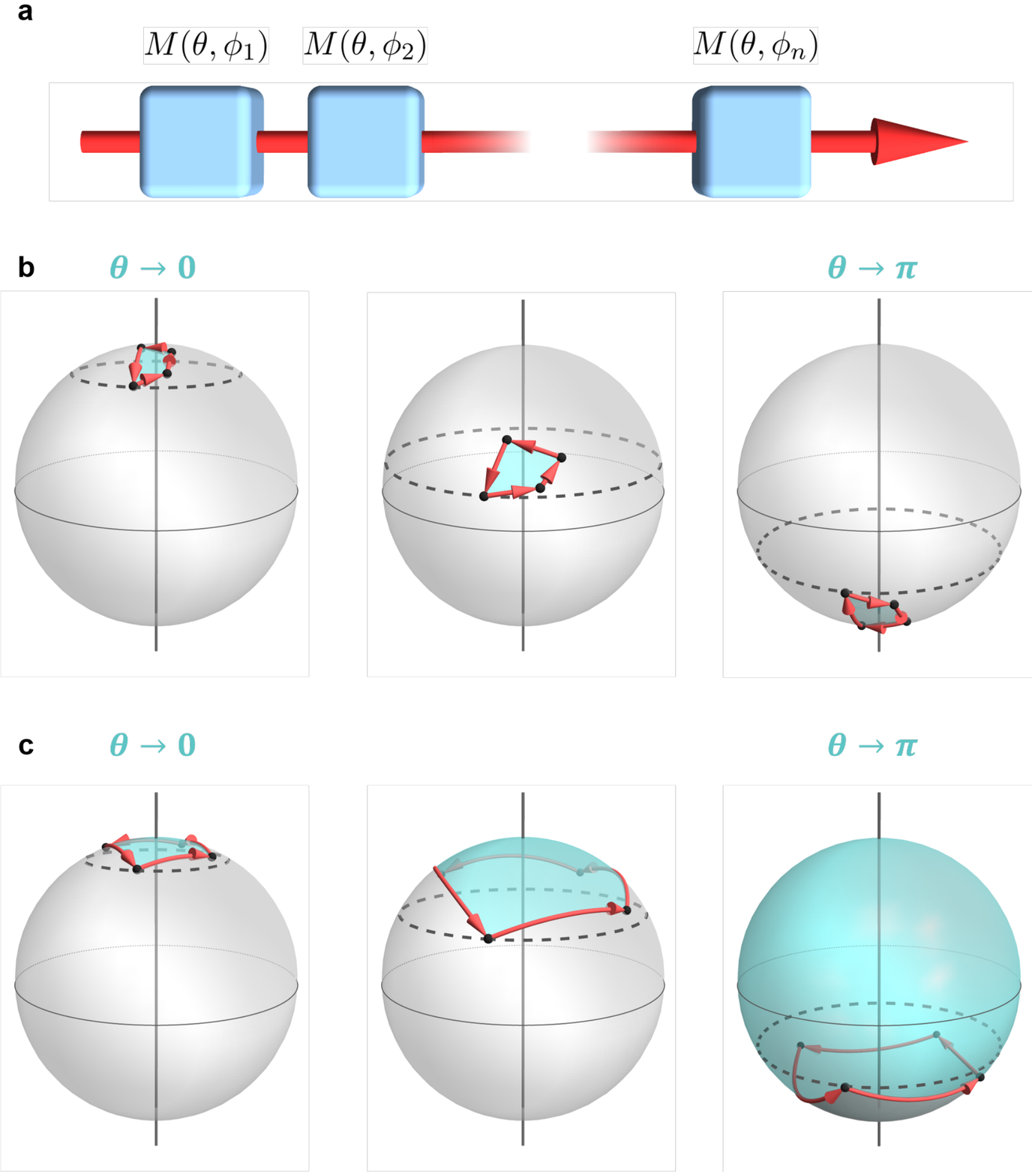}
\par\end{centering}
\caption{\label{fig:Strength}\textbf{Measurement-induced phase and its topological transition. a.} The state trajectory of a system is determined by a series of $N$ measurements along different directions $(\theta, \phi)$. \textbf{b} and \textbf{c}. The state trajectory on the Bloch sphere for a sequence of three measurements with strength $\eta=0.2$ (\textbf{b}) and $\eta=0.7$ (\textbf{c}) for different values of $\theta$. The black dashed line corresponds to $\theta$, at which the measurements are performed. The black points connected by red arrows denote the system state trajectory, as induced by the measurements. The colored portion of the Bloch sphere is the solid angle subtended by the respective trajectory.}
\end{figure}

\noindent\textbf{Theoretical overview.} We consider a class of processes where $N$ measurements are performed on a quantum system, as shown in Fig.~\ref{fig:Strength}a. Each step is a post-selected measurement associated with the polarization state $\ket{\theta, \phi}$, where $\theta$ and $\phi$ stand for the polar and azimuthal coordinates on the Bloch sphere, respectively. We can define a sequence of measurements $(\theta, \phi_n)$ for a fixed value of $\theta\in[0,\pi]$ while the azimuth is spanned in discrete steps $\phi_n=2\pi n/(N+1)$. Let us denote the acquired geometric phase $\chi_{\eta}(\theta)$, where $\eta \in [0, \infty)$ is introduced to indicate the strength of the measurement. It can be shown that $\Delta\chi_{\eta}=\chi_{\eta}(\pi)-\chi_{\eta}(0)=2\pi m$, where $m$ is an integer -- see Supplementary Information (SI), ~\ref{subsec:top-trans}, for more details.

As illustrated in Fig.~\ref{fig:Strength}b, for infinitely weak measurements, $\eta \rightarrow 0$, the effect of each measurement is vanishingly small. Therefore, $\chi(\theta)=0$ for any value of $\theta$, implying that $\Delta\chi_{\eta\rightarrow0}=0$. However, in the limit of projective measurements, strong measurement, one observes that $\Delta\chi_{\eta\rightarrow\infty}=2\pi$, as shown in Pancharatnam's geometric-phase theory~\cite{Pancharatnam1956}. An example of the latter case, strong measurement limit, is illustrated in Fig.~\ref{fig:Strength}c: When $\theta = 0$, the measurement sequence does not change the projected state. Thus, the trajectory on the Bloch sphere shrinks to a single point independently of the measurement strength. In consequence, the enclosed area -- and the geometrical phase -- is zero. For $\theta \rightarrow 0$ , the state follows a loop close to the initial projected state, acquiring a small geometric phase. As $\theta \rightarrow \pi$, the state follows a similar loop close to the south pole of the Bloch sphere, thus the enclosed geometric phase is close to $2\pi$. This gives $\Delta\chi_{\eta\rightarrow\infty} = 2\pi$, as stated above. The distinction between $\Delta\chi_{\eta\rightarrow0}=0$ and $\Delta\chi_{\eta\rightarrow\infty} = 2\pi$ suggests the existence of a transition in the behavior of the geometric phase, as the measurement strength $\eta$ is varied. Since $\Delta \chi = 2 \pi m $, the nature of the transition is topological.\newline

\begin{figure*}
\begin{centering}
\includegraphics[width=1\textwidth]{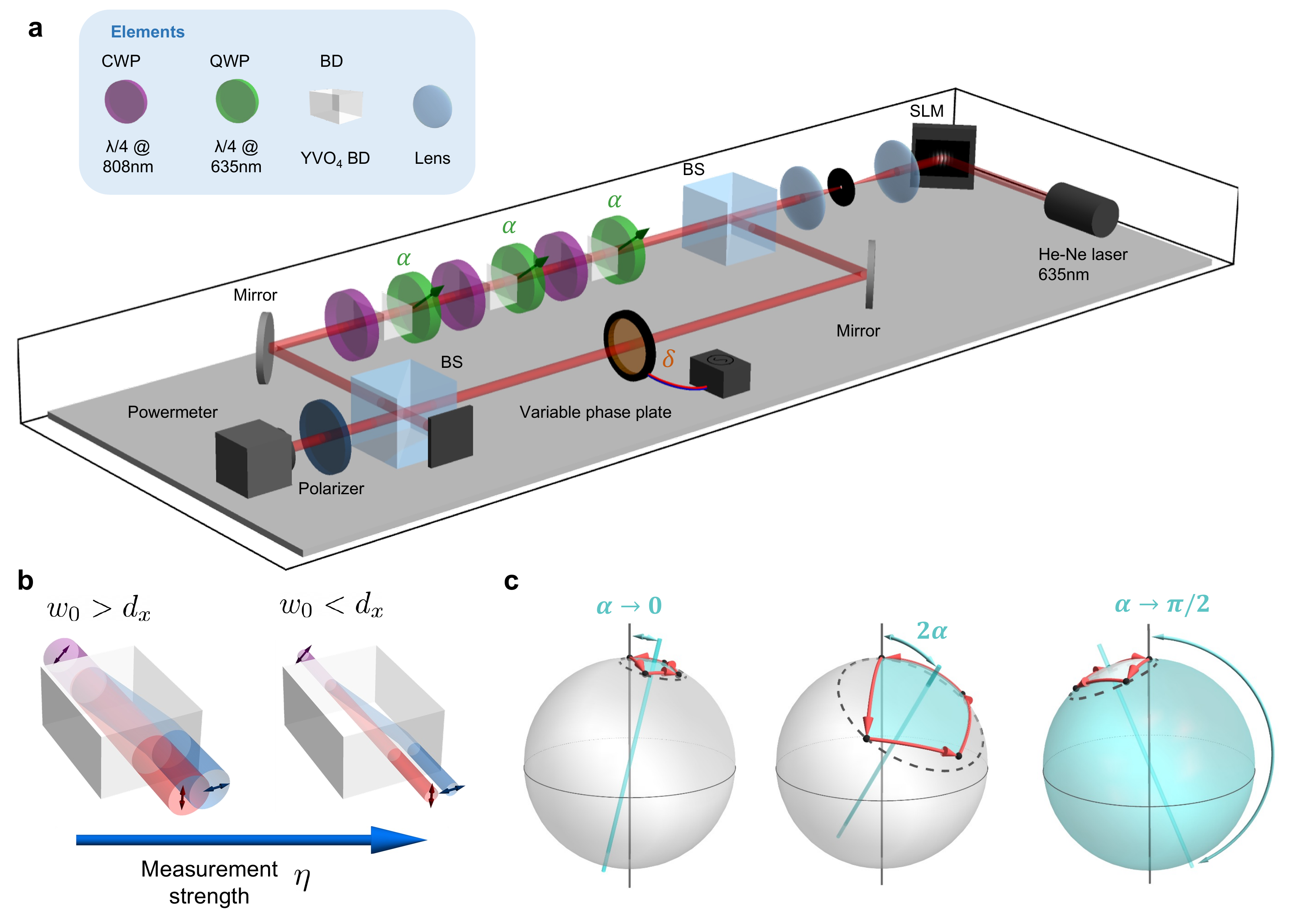}
\par\end{centering}
\caption{\label{fig:Experimental} \textbf{Optical implementation of a sequence of weak polarization measurements. a.}~Experimental setup used to detect the geometrical phase acquired due to a sequence of polarization measurements. A 632.9 nm laser emits a vertically polarized Gaussian beam which impinges on a spatial light modulator (SLM) to obtain a beam with a certain width $w_{0}$. The beam is split into two paths: in one is subjected to a sequence of transformations, while a spatially uniform phase $\delta$ is applied to the other path. Each stage is composed of a quarter-wave plate (QWP), whose fast axis is oriented at angle $\alpha$ with respect to the vertical, a beam displacer (BD), and a QWP for 808~nm acting as the compensation wave plate (CWP). Finally, the output power of the interference is recorded after the recombined beam passes through a vertical polarizer. \textbf{b.}~The measurement strength $\eta = d_x/w_0$ is controlled by varying the waist parameter $w_0$ of the input beam. When $w_0$ is much larger than the beam displacement $d_x$, the displacement is ineffectual, corresponding to a weak measurement. For $w_0 < d_x$, the two polarizations become two well-separated beams, leading to a projective measurement in the limit $w_0 \rightarrow 0$. \textbf{c.}~In contrast to the examples exhibited in Fig.~\ref{fig:Strength}b-c, the sequence of measurements produced by this setup corresponds to a circle of $\theta = 2\alpha$, which is additionally rotated by $2\alpha$ around the \textit{x} axis of the Bloch sphere. This does not affect the subtended area and, consequently, the accumulated geometric phase.}
\end{figure*}

\noindent\textbf{Experimental setup.} We demonstrate the existence of this topological transition in an optical experiment where the qubit state is associated with the polarization of a coherent beam. As illustrated in Fig.~\ref{fig:Experimental}a, the beam goes through a series of $N=3$ identical optical stages which emulate the measurement steps. Each stage is composed of a quarter-wave plate (QWP), whose fast axis is oriented at angle $\alpha=\theta/2$ with respect to the vertical, followed by a YVO$_{4}$ beam displacer (BD) and an additional compensating wave plate (CWP). The BD's ordinary and extraordinary axes are aligned along $\mathbf{\hat{y}}$ and $\mathbf{\hat{x}}$, respectively. Therefore, the BD shifts the centroid of the horizontally polarized component by a distance $d_{x}$, keeping the vertically polarized contribution unchanged. The BD essentially performs a measurement in the vertical/horizontal polarization basis, as the horizontally polarized component of the beam is spatially displaced. If the beam waist $w_0$ is larger than $d_x$, the measurement is weak, since there is no sharp separation between the two polarizations components (see Fig.~\ref{fig:Experimental}b). If the waist is much smaller than the displacement, $w_0\ll d_x$, this implements a projective measurement, as the two polarization components are completely separated. Therefore, we can control the measurement strength by modifying $w_0$. The CWP with a vertically aligned fast axis is used to compensate for the phase difference between the two polarization components accumulated while propagating inside the BD.

\begin{figure*}
\begin{centering}
\includegraphics[width=1\textwidth]{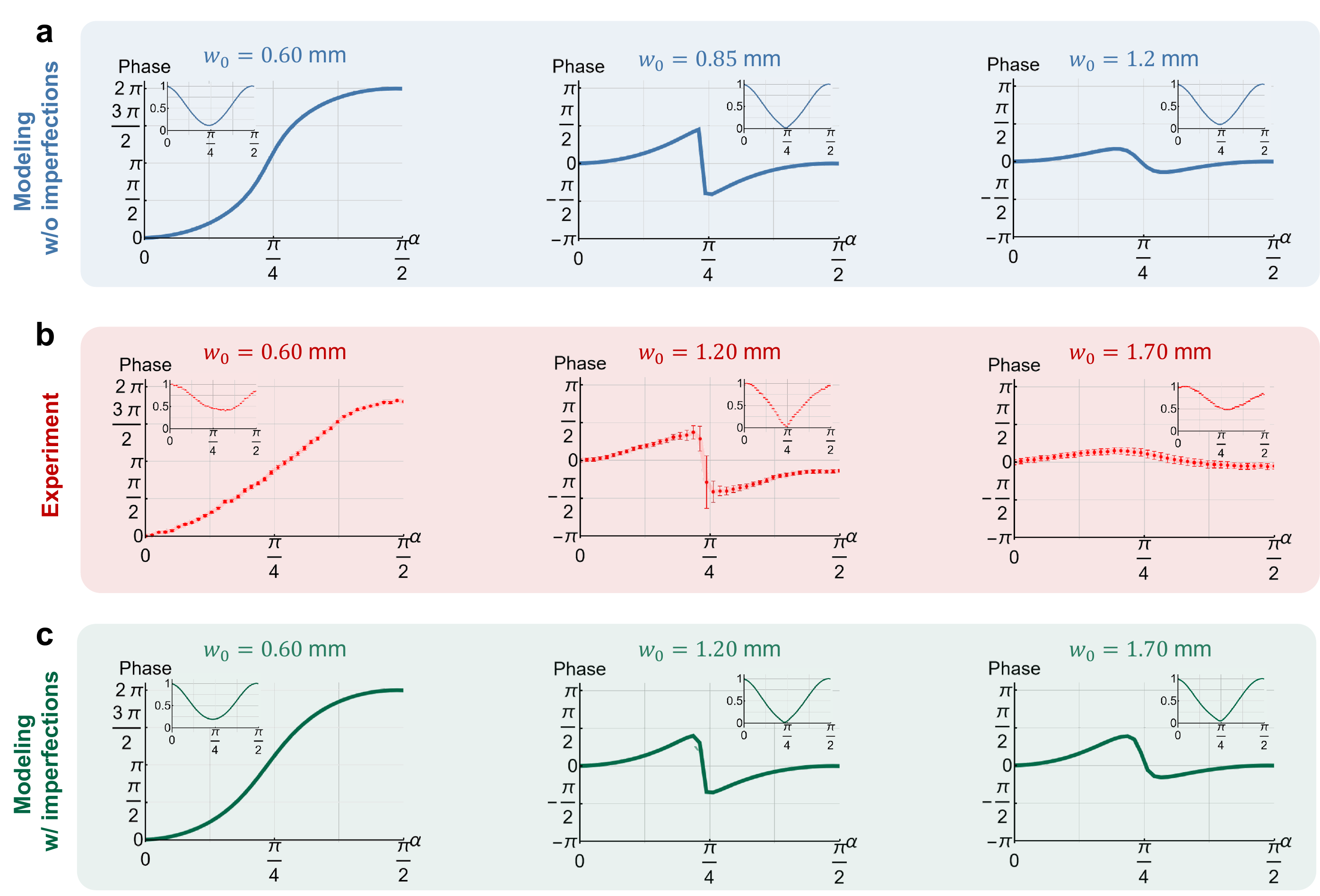}
\par\end{centering}
\caption{\label{fig:observation_results}\textbf{Experimentally measured and theoretically simulated geometric phase.} Topological transition in the measurement-induced geometric phase $\chi(\alpha = \theta/2)$: \textbf{a.} theoretical modeling, \textbf{b.} experimental results, \textbf{c.} modeling incorporating the imperfection of the birefringent crystals. The left column corresponds to a narrow beam (small $w_{0}$, strong measurement) and features $\Delta\chi = 2\pi$. The right column corresponds to a large beam width (weak measurement) and exhibits $\Delta\chi=0$. The middle column represents a point close to the transition: the phase $\chi(\alpha)$ exhibits a sharp change near $\alpha=\pi/4$, and the interference contrast vanishes near $\alpha=\pi/4$, enabling the change of the phase's topological behavior.}
\end{figure*}

The role of the QWPs is to implement the desired sequence of measurement directions $(\theta, \phi_n)$ on the Bloch sphere. The rotation by angle $\alpha$ enables controlling the polar angle $\theta$ of the measurement axis. The sequence of measurements induced by the setup in Fig.~\ref{fig:Experimental}a corresponds to the directions $(\theta, \phi_n)$ rotated by an angle $\theta = 2 \alpha$ around the \textit{y} axis of the Bloch sphere, cf.~Fig.~\ref{fig:Experimental}c.  The details of this correspondence are explained in the \ref{SI:A6}. Given the geometric nature of the induced phase, the expected topological transition remains unaffected by this rotation, both qualitatively and quantitatively. Finally, to complete the cyclic evolution, the polarization state is projected onto the initial state using a polarizer. 

Our aim is to investigate the geometrical phase acquired by the non-deflected beam (corresponding to the measurement postselected to yield a null outcome). This is done by interfering the final state with the reference beam, which only experiences a controllable phase shift $\delta$ introduced by a variable phase plate. The output power at the interferometer exit is recorded as a function of $\delta$. The shift of this curve corresponds to the acquired geometric phase. The input beam is generated by means of a spatial light modulator (SLM) that displays a hologram allowing to tailor the beam waist through the technique introduced in Ref.~\cite{bolduc2013exact}. We set the input beam's polarization state to be vertical ($\mathbf{\hat{y}}$), corresponding to the initial state in the direction $(\theta, \phi_0)$ in the theoretical protocol. Based on this setup, the measurement protocol to unveil the hidden topological transition is given as follows. The strength of our intermediate $N = 3$ measurements is regulated by varying the waist parameter of the input beam: the value of $w_{0}$ is inversely proportional to the measurement strength $\eta = d_x/w_0$ (see Fig.~\ref{fig:Experimental}c). For a fixed waist parameter, we proceed to get power readouts as a function of the reference arm phase shift $\delta\in[0,2\pi]$, while $\alpha=\theta/2$ is kept constant. From here, it is possible to retrieve the accumulated geometrical phase $\chi_{\eta}(\theta = 2\alpha)$ for a given orientation of QWPs, $\alpha$, by proper curve fitting. By varying the QWPs orientation, it is possible to reconstruct the behavior of $\chi_{\eta}(\theta = 2\alpha)$ for all $\alpha$ for a given measurement strength.\newline

\noindent\textbf{Experimental results.} Here, we discuss the experimental results and their relation to the theoretical predictions. Since the postselection in our experiment goes beyond the original theoretical proposal, we have modeled the experiment to confirm the presence of a topological transition theoretically (see  \ref{subsec:measurement_implementation} of SI). Figure~\ref{fig:observation_results}a shows that when $w_0$ is sufficiently small, i.e. strong measurement regime, the simulation predicts $\Delta\chi=\chi(\alpha = \pi/2)-\chi(\alpha = 0)=2\pi$, while for the case of weak measurements ($w_0<d_x$), $\Delta\chi=0$. A sharp transition is occurs at $w_{0}=0.85$~mm, where the interference contrast vanishes for $\alpha\approx\pi/4$, enabling the abrupt change of the phase behavior. The experiment was carried on by performing measurements for $w_{0}$ between 0.6 mm and 2.5 mm. The experimental results, shown in Fig.~\ref{fig:observation_results}b, clearly exhibit a similar transition between $\Delta\chi=2\pi$ for small $w_{0}$ and $\Delta\chi=0$ for large $w_{0}$, as well as the vanishing contrast at the transition.

The difference $\Delta\chi=\chi(\pi/2)-\chi(0)$ in the observations is not strictly equal to $0$ or $2\pi$, but can slightly deviate from these values. This is seen most prominently for $w_{0}=0.6$~mm. We attribute this to the stability of the Mach-Zehnder interferometer, in particular to a small drift in the phase between the two arms during the measurement process (which was performed in 45 minutes). We emphasize that this does not violate the topological quantization of $\Delta\chi$, but introduces an error in its extraction. In all the cases, the extracted $\Delta\chi$ is close to either $0$ or $2\pi$, making the determination of the topological index $m$ straightforward. The vanishing contrast at the transition also confirms the expected phenomenology of the topological transition.

We note that the waist $w_0^*$ at which the transition happens clearly deviates from the theory predictions: $w_{0}^*=0.85$~mm in the simulation, while $w_{0}^*=1.2$~mm in the experiment. We attribute this deviation to the fact that the surfaces of the BDs are parallel within a few tens of arcseconds, as stated by the manufacturer and verified by us independently. This tiny angle between the two surfaces induces a small transverse wavevector difference between the two components. We have incorporated this effect into our theoretical modeling, the results of which are presented in Fig.~\ref{fig:observation_results}c. With this, we are able to reproduce the change in the transition location. A detailed analysis of these imperfections and the enhanced modeling can be found in \ref{SI:Imper} of the SI.

\begin{figure}
\begin{centering}
\includegraphics[width=1\columnwidth]{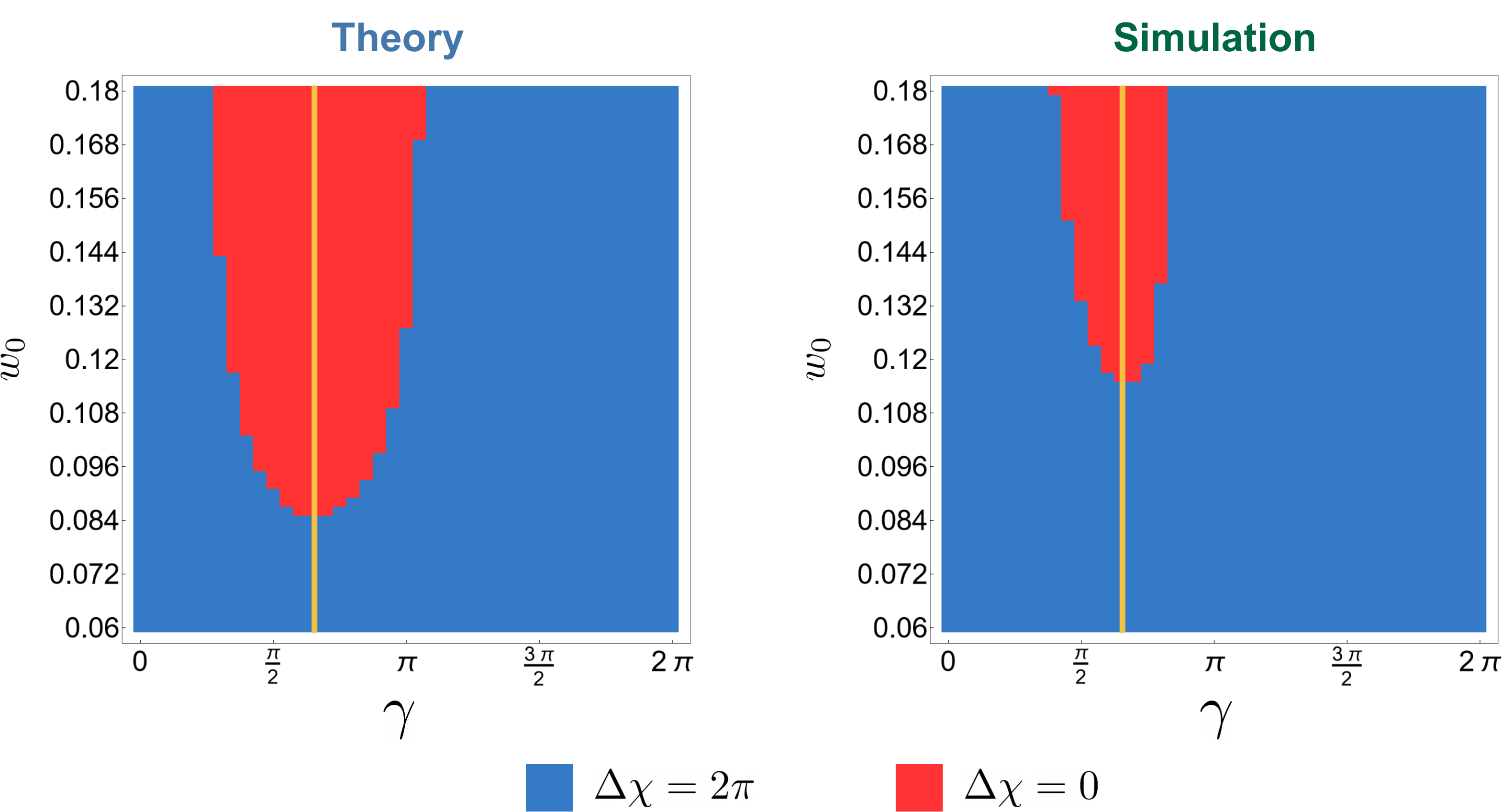}
\par\end{centering}
\caption{\label{fig:phase_diagram} \textbf{Topological phase dependence on compensating waveplates.} The phase diagram (theoretical) depicting the topological properties of the measurement-induced phase as a function of beam waist $w_{0}$ and the compensation phase $\gamma$. We present the results of a theory simulation without (left) and including (right) the experimental imperfections of the birefringent crystals. Note that the trivial phase with ($\Delta\chi=0$) exists only in a narrow interval of the phase compensation parameter 
The vertical line indicates the parameters used in our experiment. The imperfections of birefringent crystals clearly make the trivial region shrink, yet do not eliminate the topological transition.}
\end{figure}

Theoretical studies have predicted~\citep{Snizhko2021,Snizhko2021a} that the topological transition only exists if the dynamical phases are compensated accurately enough. In our work, this condition is satisfied. In Fig.~\ref{fig:phase_diagram} we explored theoretically the topological phase diagram considering the additional parameter $\gamma$ corresponding to the optical retardation of the CWP. The results show that there is a range of values of $\gamma$ where the topological transition can be observed, both in the ideal scenario and in the case of imperfect optical elements. 

\section{Discussion}
We have demonstrated that measurement-induced geometric phases in optical systems exhibit a topological transition. In particular, we consider a family of processes parameterized by a variable $\alpha$ and a measurement strength $\eta$. We demonstrated that the geometrical phase, with respect to $\eta$ and $\alpha$, exhibits a nontrivial topology. More precisely, the variation $\Delta\chi$ in geometrical phase as a function of $\alpha$ undergoes a sharp transition of $2\pi$ as $\eta$ is varied. The parameter $\eta$ can be viewed as the coupling strength with an environment, here represented by the light's spatial degree of freedom. In this framework, our observations can be interpreted as topological transitions induced by the coupling to an external environment. The topological transition is robust to amending the protocol and the imperfections in the measurement process. The location of the topological transition depends on specific details of the system (quality of beam displacers, retardation of the compensating waveplates, etc.). This sensitivity of the transition location may be useful for characterizing optical elements or for sensing. We leave this, however, to future investigations.



\bibliography{BibTTGBP.bib}

\begin{thebibliography}{28}%
\makeatletter
\providecommand \@ifxundefined [1]{%
 \@ifx{#1\undefined}
}%
\providecommand \@ifnum [1]{%
 \ifnum #1\expandafter \@firstoftwo
 \else \expandafter \@secondoftwo
 \fi
}%
\providecommand \@ifx [1]{%
 \ifx #1\expandafter \@firstoftwo
 \else \expandafter \@secondoftwo
 \fi
}%
\providecommand \natexlab [1]{#1}%
\providecommand \enquote  [1]{``#1''}%
\providecommand \bibnamefont  [1]{#1}%
\providecommand \bibfnamefont [1]{#1}%
\providecommand \citenamefont [1]{#1}%
\providecommand \href@noop [0]{\@secondoftwo}%
\providecommand \href [0]{\begingroup \@sanitize@url \@href}%
\providecommand \@href[1]{\@@startlink{#1}\@@href}%
\providecommand \@@href[1]{\endgroup#1\@@endlink}%
\providecommand \@sanitize@url [0]{\catcode `\\12\catcode `\$12\catcode
  `\&12\catcode `\#12\catcode `\^12\catcode `\_12\catcode `\%12\relax}%
\providecommand \@@startlink[1]{}%
\providecommand \@@endlink[0]{}%
\providecommand \url  [0]{\begingroup\@sanitize@url \@url }%
\providecommand \@url [1]{\endgroup\@href {#1}{\urlprefix }}%
\providecommand \urlprefix  [0]{URL }%
\providecommand \Eprint [0]{\href }%
\providecommand \doibase [0]{https://doi.org/}%
\providecommand \selectlanguage [0]{\@gobble}%
\providecommand \bibinfo  [0]{\@secondoftwo}%
\providecommand \bibfield  [0]{\@secondoftwo}%
\providecommand \translation [1]{[#1]}%
\providecommand \BibitemOpen [0]{}%
\providecommand \bibitemStop [0]{}%
\providecommand \bibitemNoStop [0]{.\EOS\space}%
\providecommand \EOS [0]{\spacefactor3000\relax}%
\providecommand \BibitemShut  [1]{\csname bibitem#1\endcsname}%
\let\auto@bib@innerbib\@empty
\bibitem [{\citenamefont {Berry}(1984)}]{Berry1984}%
  \BibitemOpen
  \bibfield  {author} {\bibinfo {author} {\bibfnamefont {M.~V.}\ \bibnamefont
  {Berry}},\ }\bibfield  {title} {\bibinfo {title} {{Quantal Phase Factors
  Accompanying Adiabatic Changes}},\ }\href
  {https://doi.org/10.1098/rspa.1984.0023} {\bibfield  {journal} {\bibinfo
  {journal} {Proceedings of the Royal Society A: Mathematical, Physical and
  Engineering Sciences}\ }\textbf {\bibinfo {volume} {392}},\ \bibinfo {pages}
  {45} (\bibinfo {year} {1984})}\BibitemShut {NoStop}%
\bibitem [{\citenamefont {Pancharatnam}(1956)}]{Pancharatnam1956}%
  \BibitemOpen
  \bibfield  {author} {\bibinfo {author} {\bibfnamefont {S.}~\bibnamefont
  {Pancharatnam}},\ }\bibfield  {title} {\bibinfo {title} {{Generalized theory
  of interference, and its applications - part I}},\ }\href
  {https://doi.org/10.1007/bf03046050} {\bibfield  {journal} {\bibinfo
  {journal} {Proceedings of the Indian Academy of Sciences - Section A}\
  }\textbf {\bibinfo {volume} {44}},\ \bibinfo {pages} {247} (\bibinfo {year}
  {1956})}\BibitemShut {NoStop}%
\bibitem [{\citenamefont {Berry}\ and\ \citenamefont
  {Klein}(1996)}]{Berry1996}%
  \BibitemOpen
  \bibfield  {author} {\bibinfo {author} {\bibfnamefont {M.~V.}\ \bibnamefont
  {Berry}}\ and\ \bibinfo {author} {\bibfnamefont {S.}~\bibnamefont {Klein}},\
  }\bibfield  {title} {\bibinfo {title} {{Geometric phases from stacks of
  crystal plates}},\ }\href
  {http://www.tandfonline.com/doi/abs/10.1080/09500349608232731} {\bibfield
  {journal} {\bibinfo  {journal} {J.Mod.Opt}\ }\textbf {\bibinfo {volume}
  {43}},\ \bibinfo {pages} {165} (\bibinfo {year} {1996})}\BibitemShut
  {NoStop}%
\bibitem [{\citenamefont {Chruscinski}\ and\ \citenamefont
  {Jamiolkowski}(2004)}]{Chruscinski2004}%
  \BibitemOpen
  \bibfield  {author} {\bibinfo {author} {\bibfnamefont {D.}~\bibnamefont
  {Chruscinski}}\ and\ \bibinfo {author} {\bibfnamefont {A.}~\bibnamefont
  {Jamiolkowski}},\ }\href@noop {} {\emph {\bibinfo {title} {{Geometric phases
  in classical and quantum mechanics}}}}\ (\bibinfo  {publisher}
  {Birkh{\"{a}}user Basel},\ \bibinfo {year} {2004})\BibitemShut {NoStop}%
\bibitem [{\citenamefont {Cohen}\ \emph {et~al.}(2019)\citenamefont {Cohen},
  \citenamefont {Larocque}, \citenamefont {Bouchard}, \citenamefont
  {Nejadsattari}, \citenamefont {Gefen},\ and\ \citenamefont
  {Karimi}}]{Cohen2019}%
  \BibitemOpen
  \bibfield  {author} {\bibinfo {author} {\bibfnamefont {E.}~\bibnamefont
  {Cohen}}, \bibinfo {author} {\bibfnamefont {H.}~\bibnamefont {Larocque}},
  \bibinfo {author} {\bibfnamefont {F.}~\bibnamefont {Bouchard}}, \bibinfo
  {author} {\bibfnamefont {F.}~\bibnamefont {Nejadsattari}}, \bibinfo {author}
  {\bibfnamefont {Y.}~\bibnamefont {Gefen}},\ and\ \bibinfo {author}
  {\bibfnamefont {E.}~\bibnamefont {Karimi}},\ }\bibfield  {title} {\bibinfo
  {title} {{Geometric phase from Aharonov–Bohm to Pancharatnam–Berry and
  beyond}},\ }\href {https://doi.org/10.1038/s42254-019-0071-1} {\bibfield
  {journal} {\bibinfo  {journal} {Nature Reviews Physics}\ }\textbf {\bibinfo
  {volume} {1}},\ \bibinfo {pages} {437} (\bibinfo {year} {2019})}\BibitemShut
  {NoStop}%
\bibitem [{\citenamefont {Bomzon}\ \emph {et~al.}(2002)\citenamefont {Bomzon},
  \citenamefont {Biener}, \citenamefont {Kleiner},\ and\ \citenamefont
  {Hasman}}]{bomzon2002space}%
  \BibitemOpen
  \bibfield  {author} {\bibinfo {author} {\bibfnamefont {Z.}~\bibnamefont
  {Bomzon}}, \bibinfo {author} {\bibfnamefont {G.}~\bibnamefont {Biener}},
  \bibinfo {author} {\bibfnamefont {V.}~\bibnamefont {Kleiner}},\ and\ \bibinfo
  {author} {\bibfnamefont {E.}~\bibnamefont {Hasman}},\ }\bibfield  {title}
  {\bibinfo {title} {Space-variant pancharatnam--berry phase optical elements
  with computer-generated subwavelength gratings},\ }\href@noop {} {\bibfield
  {journal} {\bibinfo  {journal} {Optics letters}\ }\textbf {\bibinfo {volume}
  {27}},\ \bibinfo {pages} {1141} (\bibinfo {year} {2002})}\BibitemShut
  {NoStop}%
\bibitem [{\citenamefont {Bliokh}\ \emph {et~al.}(2019)\citenamefont {Bliokh},
  \citenamefont {Alonso},\ and\ \citenamefont {Dennis}}]{bliokh2019geometric}%
  \BibitemOpen
  \bibfield  {author} {\bibinfo {author} {\bibfnamefont {K.~Y.}\ \bibnamefont
  {Bliokh}}, \bibinfo {author} {\bibfnamefont {M.~A.}\ \bibnamefont {Alonso}},\
  and\ \bibinfo {author} {\bibfnamefont {M.~R.}\ \bibnamefont {Dennis}},\
  }\bibfield  {title} {\bibinfo {title} {Geometric phases in 2d and 3d
  polarized fields: geometrical, dynamical, and topological aspects},\
  }\href@noop {} {\bibfield  {journal} {\bibinfo  {journal} {Reports on
  Progress in Physics}\ }\textbf {\bibinfo {volume} {82}},\ \bibinfo {pages}
  {122401} (\bibinfo {year} {2019})}\BibitemShut {NoStop}%
\bibitem [{\citenamefont {Rubano}\ \emph {et~al.}(2019)\citenamefont {Rubano},
  \citenamefont {Cardano}, \citenamefont {Piccirillo},\ and\ \citenamefont
  {Marrucci}}]{rubano2019q}%
  \BibitemOpen
  \bibfield  {author} {\bibinfo {author} {\bibfnamefont {A.}~\bibnamefont
  {Rubano}}, \bibinfo {author} {\bibfnamefont {F.}~\bibnamefont {Cardano}},
  \bibinfo {author} {\bibfnamefont {B.}~\bibnamefont {Piccirillo}},\ and\
  \bibinfo {author} {\bibfnamefont {L.}~\bibnamefont {Marrucci}},\ }\bibfield
  {title} {\bibinfo {title} {Q-plate technology: a progress review},\
  }\href@noop {} {\bibfield  {journal} {\bibinfo  {journal} {JOSA B}\ }\textbf
  {\bibinfo {volume} {36}},\ \bibinfo {pages} {D70} (\bibinfo {year}
  {2019})}\BibitemShut {NoStop}%
\bibitem [{\citenamefont {Jisha}\ \emph {et~al.}(2021)\citenamefont {Jisha},
  \citenamefont {Nolte},\ and\ \citenamefont {Alberucci}}]{jisha2021geometric}%
  \BibitemOpen
  \bibfield  {author} {\bibinfo {author} {\bibfnamefont {C.~P.}\ \bibnamefont
  {Jisha}}, \bibinfo {author} {\bibfnamefont {S.}~\bibnamefont {Nolte}},\ and\
  \bibinfo {author} {\bibfnamefont {A.}~\bibnamefont {Alberucci}},\ }\bibfield
  {title} {\bibinfo {title} {Geometric phase in optics: from wavefront
  manipulation to waveguiding},\ }\href@noop {} {\bibfield  {journal} {\bibinfo
   {journal} {Laser \& Photonics Reviews}\ }\textbf {\bibinfo {volume} {15}},\
  \bibinfo {pages} {2100003} (\bibinfo {year} {2021})}\BibitemShut {NoStop}%
\bibitem [{\citenamefont {Longuet-Higgins}\ \emph {et~al.}(1958)\citenamefont
  {Longuet-Higgins}, \citenamefont {Öpik}, \citenamefont {Pryce},\ and\
  \citenamefont {Sack}}]{LH:58}%
  \BibitemOpen
  \bibfield  {author} {\bibinfo {author} {\bibfnamefont {H.~C.}\ \bibnamefont
  {Longuet-Higgins}}, \bibinfo {author} {\bibfnamefont {U.}~\bibnamefont
  {Öpik}}, \bibinfo {author} {\bibfnamefont {M.~H.~L.}\ \bibnamefont
  {Pryce}},\ and\ \bibinfo {author} {\bibfnamefont {R.~A.}\ \bibnamefont
  {Sack}},\ }\bibfield  {title} {\bibinfo {title} {Studies of the jahn-teller
  effect .ii. the dynamical problem},\ }\href
  {https://doi.org/10.1098/rspa.1958.0022} {\bibfield  {journal} {\bibinfo
  {journal} {Proceedings of the Royal Society of London. Series A. Mathematical
  and Physical Sciences}\ }\textbf {\bibinfo {volume} {244}},\ \bibinfo {pages}
  {1} (\bibinfo {year} {1958})}\BibitemShut {NoStop}%
\bibitem [{\citenamefont {Zak}(1989)}]{zak1989berry}%
  \BibitemOpen
  \bibfield  {author} {\bibinfo {author} {\bibfnamefont {J.}~\bibnamefont
  {Zak}},\ }\bibfield  {title} {\bibinfo {title} {Berry’s phase for energy
  bands in solids},\ }\href@noop {} {\bibfield  {journal} {\bibinfo  {journal}
  {Physical review letters}\ }\textbf {\bibinfo {volume} {62}},\ \bibinfo
  {pages} {2747} (\bibinfo {year} {1989})}\BibitemShut {NoStop}%
\bibitem [{\citenamefont {Resta}(2000)}]{resta2000manifestations}%
  \BibitemOpen
  \bibfield  {author} {\bibinfo {author} {\bibfnamefont {R.}~\bibnamefont
  {Resta}},\ }\bibfield  {title} {\bibinfo {title} {Manifestations of berry's
  phase in molecules and condensed matter},\ }\href@noop {} {\bibfield
  {journal} {\bibinfo  {journal} {Journal of Physics: Condensed Matter}\
  }\textbf {\bibinfo {volume} {12}},\ \bibinfo {pages} {R107} (\bibinfo {year}
  {2000})}\BibitemShut {NoStop}%
\bibitem [{\citenamefont {Nayak}\ \emph {et~al.}(2008)\citenamefont {Nayak},
  \citenamefont {Simon}, \citenamefont {Stern}, \citenamefont {Freedman},\ and\
  \citenamefont {Sarma}}]{nayak2008non}%
  \BibitemOpen
  \bibfield  {author} {\bibinfo {author} {\bibfnamefont {C.}~\bibnamefont
  {Nayak}}, \bibinfo {author} {\bibfnamefont {S.~H.}\ \bibnamefont {Simon}},
  \bibinfo {author} {\bibfnamefont {A.}~\bibnamefont {Stern}}, \bibinfo
  {author} {\bibfnamefont {M.}~\bibnamefont {Freedman}},\ and\ \bibinfo
  {author} {\bibfnamefont {S.~D.}\ \bibnamefont {Sarma}},\ }\bibfield  {title}
  {\bibinfo {title} {Non-abelian anyons and topological quantum computation},\
  }\href@noop {} {\bibfield  {journal} {\bibinfo  {journal} {Reviews of Modern
  Physics}\ }\textbf {\bibinfo {volume} {80}},\ \bibinfo {pages} {1083}
  (\bibinfo {year} {2008})}\BibitemShut {NoStop}%
\bibitem [{\citenamefont {Fradkin}(2013)}]{fradkin2013field}%
  \BibitemOpen
  \bibfield  {author} {\bibinfo {author} {\bibfnamefont {E.}~\bibnamefont
  {Fradkin}},\ }\href@noop {} {\emph {\bibinfo {title} {Field theories of
  condensed matter physics}}}\ (\bibinfo  {publisher} {Cambridge University
  Press},\ \bibinfo {year} {2013})\BibitemShut {NoStop}%
\bibitem [{\citenamefont {Resta}(1994)}]{resta1994macroscopic}%
  \BibitemOpen
  \bibfield  {author} {\bibinfo {author} {\bibfnamefont {R.}~\bibnamefont
  {Resta}},\ }\bibfield  {title} {\bibinfo {title} {Macroscopic polarization in
  crystalline dielectrics: the geometric phase approach},\ }\href@noop {}
  {\bibfield  {journal} {\bibinfo  {journal} {Reviews of modern physics}\
  }\textbf {\bibinfo {volume} {66}},\ \bibinfo {pages} {899} (\bibinfo {year}
  {1994})}\BibitemShut {NoStop}%
\bibitem [{\citenamefont {Asb{\'o}th}\ \emph {et~al.}(2016)\citenamefont
  {Asb{\'o}th}, \citenamefont {Oroszl{\'a}ny},\ and\ \citenamefont
  {P{\'a}lyi}}]{asboth2016short}%
  \BibitemOpen
  \bibfield  {author} {\bibinfo {author} {\bibfnamefont {J.~K.}\ \bibnamefont
  {Asb{\'o}th}}, \bibinfo {author} {\bibfnamefont {L.}~\bibnamefont
  {Oroszl{\'a}ny}},\ and\ \bibinfo {author} {\bibfnamefont {A.}~\bibnamefont
  {P{\'a}lyi}},\ }\bibfield  {title} {\bibinfo {title} {A short course on
  topological insulators},\ }\href@noop {} {\bibfield  {journal} {\bibinfo
  {journal} {Lecture notes in physics}\ }\textbf {\bibinfo {volume} {919}},\
  \bibinfo {pages} {166} (\bibinfo {year} {2016})}\BibitemShut {NoStop}%
\bibitem [{\citenamefont {Hasan}\ and\ \citenamefont
  {Kane}(2010)}]{hasan2010colloquium}%
  \BibitemOpen
  \bibfield  {author} {\bibinfo {author} {\bibfnamefont {M.~Z.}\ \bibnamefont
  {Hasan}}\ and\ \bibinfo {author} {\bibfnamefont {C.~L.}\ \bibnamefont
  {Kane}},\ }\bibfield  {title} {\bibinfo {title} {Colloquium: topological
  insulators},\ }\href@noop {} {\bibfield  {journal} {\bibinfo  {journal}
  {Reviews of modern physics}\ }\textbf {\bibinfo {volume} {82}},\ \bibinfo
  {pages} {3045} (\bibinfo {year} {2010})}\BibitemShut {NoStop}%
\bibitem [{\citenamefont {Garrison}\ and\ \citenamefont
  {Wright}(1988)}]{garrison1988complex}%
  \BibitemOpen
  \bibfield  {author} {\bibinfo {author} {\bibfnamefont {J.}~\bibnamefont
  {Garrison}}\ and\ \bibinfo {author} {\bibfnamefont {E.}~\bibnamefont
  {Wright}},\ }\bibfield  {title} {\bibinfo {title} {Complex geometrical phases
  for dissipative systems},\ }\href@noop {} {\bibfield  {journal} {\bibinfo
  {journal} {Physics Letters A}\ }\textbf {\bibinfo {volume} {128}},\ \bibinfo
  {pages} {177} (\bibinfo {year} {1988})}\BibitemShut {NoStop}%
\bibitem [{\citenamefont {Dattoli}\ \emph {et~al.}(1990)\citenamefont
  {Dattoli}, \citenamefont {Mignani},\ and\ \citenamefont
  {Torre}}]{dattoli1990geometrical}%
  \BibitemOpen
  \bibfield  {author} {\bibinfo {author} {\bibfnamefont {G.}~\bibnamefont
  {Dattoli}}, \bibinfo {author} {\bibfnamefont {R.}~\bibnamefont {Mignani}},\
  and\ \bibinfo {author} {\bibfnamefont {A.}~\bibnamefont {Torre}},\ }\bibfield
   {title} {\bibinfo {title} {Geometrical phase in the cyclic evolution of
  non-hermitian systems},\ }\href@noop {} {\bibfield  {journal} {\bibinfo
  {journal} {Journal of Physics A: Mathematical and General}\ }\textbf
  {\bibinfo {volume} {23}},\ \bibinfo {pages} {5795} (\bibinfo {year}
  {1990})}\BibitemShut {NoStop}%
\bibitem [{\citenamefont {El-Ganainy}\ \emph {et~al.}(2018)\citenamefont
  {El-Ganainy}, \citenamefont {Makris}, \citenamefont {Khajavikhan},
  \citenamefont {Musslimani}, \citenamefont {Rotter},\ and\ \citenamefont
  {Christodoulides}}]{el2018non}%
  \BibitemOpen
  \bibfield  {author} {\bibinfo {author} {\bibfnamefont {R.}~\bibnamefont
  {El-Ganainy}}, \bibinfo {author} {\bibfnamefont {K.~G.}\ \bibnamefont
  {Makris}}, \bibinfo {author} {\bibfnamefont {M.}~\bibnamefont {Khajavikhan}},
  \bibinfo {author} {\bibfnamefont {Z.~H.}\ \bibnamefont {Musslimani}},
  \bibinfo {author} {\bibfnamefont {S.}~\bibnamefont {Rotter}},\ and\ \bibinfo
  {author} {\bibfnamefont {D.~N.}\ \bibnamefont {Christodoulides}},\ }\bibfield
   {title} {\bibinfo {title} {Non-hermitian physics and pt symmetry},\
  }\href@noop {} {\bibfield  {journal} {\bibinfo  {journal} {Nature Physics}\
  }\textbf {\bibinfo {volume} {14}},\ \bibinfo {pages} {11} (\bibinfo {year}
  {2018})}\BibitemShut {NoStop}%
\bibitem [{\citenamefont {Cho}\ \emph {et~al.}(2019)\citenamefont {Cho},
  \citenamefont {Kim}, \citenamefont {Choi}, \citenamefont {Kim}, \citenamefont
  {Han}, \citenamefont {Lee}, \citenamefont {Moon},\ and\ \citenamefont
  {Kim}}]{Cho2019}%
  \BibitemOpen
  \bibfield  {author} {\bibinfo {author} {\bibfnamefont {Y.-W.}\ \bibnamefont
  {Cho}}, \bibinfo {author} {\bibfnamefont {Y.}~\bibnamefont {Kim}}, \bibinfo
  {author} {\bibfnamefont {Y.-H.}\ \bibnamefont {Choi}}, \bibinfo {author}
  {\bibfnamefont {Y.-S.}\ \bibnamefont {Kim}}, \bibinfo {author} {\bibfnamefont
  {S.-W.}\ \bibnamefont {Han}}, \bibinfo {author} {\bibfnamefont {S.-Y.}\
  \bibnamefont {Lee}}, \bibinfo {author} {\bibfnamefont {S.}~\bibnamefont
  {Moon}},\ and\ \bibinfo {author} {\bibfnamefont {Y.-H.}\ \bibnamefont
  {Kim}},\ }\bibfield  {title} {\bibinfo {title} {{Emergence of the geometric
  phase from quantum measurement back-action}},\ }\href
  {https://doi.org/10.1038/s41567-019-0482-z} {\bibfield  {journal} {\bibinfo
  {journal} {Nature Physics}\ ,\ \bibinfo {pages} {1}} (\bibinfo {year}
  {2019})}\BibitemShut {NoStop}%
\bibitem [{\citenamefont {Gebhart}\ \emph {et~al.}(2020)\citenamefont
  {Gebhart}, \citenamefont {Snizhko}, \citenamefont {Wellens}, \citenamefont
  {Buchleitner}, \citenamefont {Romito},\ and\ \citenamefont
  {Gefen}}]{Gebhart2020}%
  \BibitemOpen
  \bibfield  {author} {\bibinfo {author} {\bibfnamefont {V.}~\bibnamefont
  {Gebhart}}, \bibinfo {author} {\bibfnamefont {K.}~\bibnamefont {Snizhko}},
  \bibinfo {author} {\bibfnamefont {T.}~\bibnamefont {Wellens}}, \bibinfo
  {author} {\bibfnamefont {A.}~\bibnamefont {Buchleitner}}, \bibinfo {author}
  {\bibfnamefont {A.}~\bibnamefont {Romito}},\ and\ \bibinfo {author}
  {\bibfnamefont {Y.}~\bibnamefont {Gefen}},\ }\bibfield  {title} {\bibinfo
  {title} {{Topological transition in measurement-induced geometric phases}},\
  }\href {https://doi.org/10.1073/pnas.1911620117} {\bibfield  {journal}
  {\bibinfo  {journal} {Proceedings of the National Academy of Sciences}\
  }\textbf {\bibinfo {volume} {117}},\ \bibinfo {pages} {5706} (\bibinfo {year}
  {2020})},\ \Eprint {https://arxiv.org/abs/1905.01147} {1905.01147}
  \BibitemShut {NoStop}%
\bibitem [{\citenamefont {Snizhko}\ \emph
  {et~al.}(2021{\natexlab{a}})\citenamefont {Snizhko}, \citenamefont {Kumar},
  \citenamefont {Rao},\ and\ \citenamefont {Gefen}}]{Snizhko2021}%
  \BibitemOpen
  \bibfield  {author} {\bibinfo {author} {\bibfnamefont {K.}~\bibnamefont
  {Snizhko}}, \bibinfo {author} {\bibfnamefont {P.}~\bibnamefont {Kumar}},
  \bibinfo {author} {\bibfnamefont {N.}~\bibnamefont {Rao}},\ and\ \bibinfo
  {author} {\bibfnamefont {Y.}~\bibnamefont {Gefen}},\ }\bibfield  {title}
  {\bibinfo {title} {{Weak-Measurement-Induced Asymmetric Dephasing:
  Manifestation of Intrinsic Measurement Chirality}},\ }\href
  {https://doi.org/10.1103/PhysRevLett.127.170401} {\bibfield  {journal}
  {\bibinfo  {journal} {Physical Review Letters}\ }\textbf {\bibinfo {volume}
  {127}},\ \bibinfo {pages} {170401} (\bibinfo {year} {2021}{\natexlab{a}})},\
  \Eprint {https://arxiv.org/abs/2006.13244} {2006.13244} \BibitemShut
  {NoStop}%
\bibitem [{\citenamefont {Snizhko}\ \emph
  {et~al.}(2021{\natexlab{b}})\citenamefont {Snizhko}, \citenamefont {Rao},
  \citenamefont {Kumar},\ and\ \citenamefont {Gefen}}]{Snizhko2021a}%
  \BibitemOpen
  \bibfield  {author} {\bibinfo {author} {\bibfnamefont {K.}~\bibnamefont
  {Snizhko}}, \bibinfo {author} {\bibfnamefont {N.}~\bibnamefont {Rao}},
  \bibinfo {author} {\bibfnamefont {P.}~\bibnamefont {Kumar}},\ and\ \bibinfo
  {author} {\bibfnamefont {Y.}~\bibnamefont {Gefen}},\ }\bibfield  {title}
  {\bibinfo {title} {{Weak-measurement-induced phases and dephasing: Broken
  symmetry of the geometric phase}},\ }\href
  {https://doi.org/10.1103/PhysRevResearch.3.043045} {\bibfield  {journal}
  {\bibinfo  {journal} {Physical Review Research}\ }\textbf {\bibinfo {volume}
  {3}},\ \bibinfo {pages} {043045} (\bibinfo {year} {2021}{\natexlab{b}})},\
  \Eprint {https://arxiv.org/abs/2006.14641} {arXiv:2006.14641} \BibitemShut
  {NoStop}%
\bibitem [{\citenamefont {Wang}\ \emph {et~al.}(2022)\citenamefont {Wang},
  \citenamefont {Snizhko}, \citenamefont {Romito}, \citenamefont {Gefen},\ and\
  \citenamefont {Murch}}]{Wang2021}%
  \BibitemOpen
  \bibfield  {author} {\bibinfo {author} {\bibfnamefont {Y.}~\bibnamefont
  {Wang}}, \bibinfo {author} {\bibfnamefont {K.}~\bibnamefont {Snizhko}},
  \bibinfo {author} {\bibfnamefont {A.}~\bibnamefont {Romito}}, \bibinfo
  {author} {\bibfnamefont {Y.}~\bibnamefont {Gefen}},\ and\ \bibinfo {author}
  {\bibfnamefont {K.}~\bibnamefont {Murch}},\ }\bibfield  {title} {\bibinfo
  {title} {{Observing a topological transition in weak-measurement-induced
  geometric phases}},\ }\href
  {https://doi.org/10.1103/PhysRevResearch.4.023179} {\bibfield  {journal}
  {\bibinfo  {journal} {Physical Review Research}\ }\textbf {\bibinfo {volume}
  {4}},\ \bibinfo {pages} {023179} (\bibinfo {year} {2022})},\ \Eprint
  {https://arxiv.org/abs/2102.05660} {2102.05660} \BibitemShut {NoStop}%
\bibitem [{\citenamefont {Bolduc}\ \emph {et~al.}(2013)\citenamefont {Bolduc},
  \citenamefont {Bent}, \citenamefont {Santamato}, \citenamefont {Karimi},\
  and\ \citenamefont {Boyd}}]{bolduc2013exact}%
  \BibitemOpen
  \bibfield  {author} {\bibinfo {author} {\bibfnamefont {E.}~\bibnamefont
  {Bolduc}}, \bibinfo {author} {\bibfnamefont {N.}~\bibnamefont {Bent}},
  \bibinfo {author} {\bibfnamefont {E.}~\bibnamefont {Santamato}}, \bibinfo
  {author} {\bibfnamefont {E.}~\bibnamefont {Karimi}},\ and\ \bibinfo {author}
  {\bibfnamefont {R.~W.}\ \bibnamefont {Boyd}},\ }\bibfield  {title} {\bibinfo
  {title} {Exact solution to simultaneous intensity and phase encryption with a
  single phase-only hologram},\ }\href {https://doi.org/10.1364/OL.38.003546}
  {\bibfield  {journal} {\bibinfo  {journal} {Opt. Lett.}\ }\textbf {\bibinfo
  {volume} {38}},\ \bibinfo {pages} {3546} (\bibinfo {year}
  {2013})}\BibitemShut {NoStop}%
\bibitem [{\citenamefont {Jacobs}(2014)}]{Jacobs2014a}%
  \BibitemOpen
  \bibfield  {author} {\bibinfo {author} {\bibfnamefont {K.}~\bibnamefont
  {Jacobs}},\ }\href {https://doi.org/10.1017/CBO9781139179027} {\emph
  {\bibinfo {title} {{Quantum Measurement Theory and its Applications}}}}\
  (\bibinfo  {publisher} {Cambridge University Press},\ \bibinfo {address}
  {Cambridge},\ \bibinfo {year} {2014})\BibitemShut {NoStop}%
\bibitem [{\citenamefont {Coley}(1999)}]{coley1999GE}%
  \BibitemOpen
  \bibfield  {author} {\bibinfo {author} {\bibfnamefont {D.~A.}\ \bibnamefont
  {Coley}},\ }\href@noop {} {\emph {\bibinfo {title} {An introduction to
  genetic algorithms for scientists and engineers}}}\ (\bibinfo  {publisher}
  {World Scientific Publishing Company},\ \bibinfo {year} {1999})\BibitemShut
  {NoStop}%
\end{thebibliography}%
\vspace{0.5cm}
\noindent\textbf{Supplementary Information} accompanies this manuscript.
\vspace{1 EM}

\noindent\textbf{Acknowledgments}
\noindent This work was supported by Canada Research Chairs (CRC), Canada First Research Excellence Fund (CFREF) Program, NRC-uOttawa Joint Centre for Extreme Quantum Photonics (JCEP) via High Throughput and Secure Networks Challenge Program at the National Research Council of Canada, Deutsche Forschungsgemeinschaft (German Research Foundation) through Project No.~277101999, TRR 183 (Project C01), and Projects No. EG 96/13-1, GO 1405/6-1, and MI 658/10-2, Helmholtz International Fellow Award, the Israeli Science Foundation (ISF), NSF Grant No. DMR-2037654, and the U.S.–Israel Binational Science Foundation (BSF). 
\vspace{1 EM}

\noindent\textbf{Author Contributions}
K.S., A.R., Y.G., and E.K. conceived the idea; M.F., A.D., K.S., and E.K. designed the experiment; M.F. and K.S. performed the theoretical simulations; M.F. and A.D. performed the experiment and collected the data; M.F.,  A.D., and K.S. analysed the data; M.F., K.S., A.D., and A.R. prepared the first version of the manuscript. All authors discussed the results and contributed to the text of the manuscript.
\vspace{1 EM}

\noindent\textbf{Author Information}
\noindent The authors declare no competing financial interests. Correspondence and requests for materials should be addressed to aderrico@uottawa.ca.
\clearpage
\onecolumngrid
\renewcommand{\figurename}{\textbf{Figure}}
\setcounter{figure}{0} \renewcommand{\thefigure}{\textbf{S{\arabic{figure}}}}
\setcounter{table}{0} \renewcommand{\thetable}{S\arabic{table}}
\setcounter{section}{0} \renewcommand{\thesection}{Section~\arabic{section}}
\setcounter{equation}{0} \renewcommand{\theequation}{S\arabic{equation}}
\onecolumngrid

\begin{center}
{\Large Supplementary Information for: \\ Topological transitions of the generalized Pancharatnam-Berry phase}
\end{center}
\vspace{1 EM}

\section{Theoretical protocol and its relation to the experimental setup}

Here we provide a brief theoretical background on quantum measurement
theory, on measurement induced geometric phases, and the topological
transition in them, as well as connect the quantum measurement formalism
to the optical setup employed in the paper.

\subsection{\label{subsec:measurement}Null-weak measurements of different observables
in the polarization space}

The formalism applies to the transition reported in the main text
when regarding the polarization of the laser beam as a quantum polarization
state $\ket{\theta,\phi}=\cos(\theta/2)\ket{\uparrow}+e^{i\phi}\sin(\theta/2)\ket{\downarrow}$
of a photon, where $\ket{\uparrow}$ and $\ket{\downarrow}$ label
the linearly independent vertical and horizontal polarizations respectively
with $\theta\in[0,\pi]$ and $\phi\in[0,2\pi)$.

In the most general setting, a measurement of a quantum system in
a state $\ket{\psi}$ returns an outcome $r$ with probability $P(r)=\langle\psi\vert M_{r}^{\dagger}M_{r}\vert\psi\rangle$,
while the state is updated as $\ket{\psi}\to\ket{\psi'}=M_{r}\ket{\psi}/\sqrt{P(r)}$.
The process is controlled by the Kraus operators $M_{r}$, which depend
on the specific detection process and fulfil $\sum_{r}M_{r}^{\dagger}M_{r}=1$
due to overall probability conservation~\citep{Jacobs2014a}. For
what we are concerned here, we specialize in a measurement process,
known as null weak measurement, with two possible outcomes, $r=+,\,-$,
and corresponding Kraus operators
\begin{equation}
M_{+}=\sqrt{\zeta}\ket{\downarrow}\bra{\downarrow},\quad M_{-}=\ket{\uparrow}\bra{\uparrow}+\sqrt{1-\zeta}\ket{\downarrow}\bra{\downarrow},\label{eq:partial}
\end{equation}
where $0\leq\zeta\leq1$ corresponds to the measurement strength $\eta=\sqrt{-\ln\left(1-\zeta\right)}$.
For $\zeta=1$ ($\eta\rightarrow\infty$), the measurement is projective:
the operation projects the state on $\ket{\downarrow}$ if $r=+$
and $\ket{\uparrow}$ if $r=-$. For $\zeta<1$, however, we either
have a collapse to $\ket{\downarrow}$ (``click'', described by
$M_{+}$) or no collapse (``no-click'' or ``null measurement'',
described by $M_{-}$). In the latter case, the system's state is
updated to a the post-measurement state, which depends on the pre-measurement
one.

The measurement procedure in Eq.~\eqref{eq:partial} corresponds
to an imperfect, or weak, measurement of $\sigma_{z}$. This can be
generalized to measure arbitrary observables, i.e. to distinguish
two arbitrary basis states. A direction $\mathbf{n}=(\sin\theta\cos\phi,\sin\theta\sin\phi,\cos\theta)$
is identified by the polar and azimuthal angles $\theta$ and $\phi$
on the Bloch sphere. The corresponding orthogonal basis states, $\ket{\uparrow_{\mathbf{n}}}$ and 
$\ket{\downarrow_{\mathbf{n}}}$, are defined as eigenstates of $\mathbf{n}\cdot\bm{\sigma}$
associated with the respective eigenvalue $\pm1$. For $\phi=0$ and
$\pi$ these states correspond to two mutually orthogonal linear polarizations,
whereas for all other values of $\phi$ they correspond to general
elliptic polarizations. The analog of Eq.~\eqref{eq:partial} is
then given by $M_{+}\left(\mathbf{n}\right)=\sqrt{\zeta}\ket{\downarrow_{\mathbf{n}}}\bra{\downarrow_{\mathbf{n}}}$,
$M_{-}\left(\mathbf{n}\right)=\ket{\uparrow_{\mathbf{n}}}\bra{\uparrow_{\mathbf{n}}}+\sqrt{1-\zeta}\ket{\downarrow_{\mathbf{n}}}\bra{\downarrow_{\mathbf{n}}}$.

\subsection{\label{subsec:phases}Measurement-induced geometric phases}

In order to induce a geometric phase, one needs to use multiple measurements.
We label them by $j=1,...,N$. Each measurement is determined by the
observable it measures, i.e., by the direction $\mathbf{n}_{j}$.
The respective Kraus operators are $M_{r_{j}}^{(j)}=M_{r_{j}}\left(\mathbf{n}_{j}\right)$,
where $r_{j}=\pm$ is the readout of measurement $j$.

For each such measurement with a given outcome, the phase of the post-measurement
state is gauge-dependent, hence non-physical. However, if a sequence
of measurements leads to a post-measurement final state $\ket{\psi_{f}}$
which is proportional to the initial one, $\ket{\psi_{0}}$, the phase
difference between $\ket{\psi_{f}}$ and $\ket{\psi_{0}}$ is a legit
observable given by
\begin{equation}
\chi_{\{r_{j}\}}={\rm arg}\sp{\psi_{0}}{\psi_{f}}={\rm arg}\bra{\psi_{0}}M_{r_{N}}^{(N)}\dots M_{r_{2}}^{(2)}M_{r_{1}}^{(1)}\ket{\psi_{0}},\label{eq:phase}
\end{equation}
where $r_{j}$ is the outcome (readout) of the $j$-th measurement,
whose effect is encoded in the Kraus operator $M_{r_{j}}^{(j)}$.

Note that even if the final state, $\ket{\psi_{f}}$, differs from
the initial one, the phase defined in Eq.~(\ref{eq:phase}) is still
well-defined: this can be understood introducing a fake projective
measurement onto the initial state, $\ket{\psi_{0}}\bra{\psi_{0}}$,
after the application of all $M_{r_{j}}^{(j)}$, in order to force
$\ket{\psi_{f}}\propto\ket{\psi_{0}}$. Therefore, Eq.~(\ref{eq:phase})
defines a legitimate observable for a general sequence of measurements.
The sequence of (normalized) post-measurement intermediate states,
$\ket{\psi_{0}}$, $M_{r_{1}}^{(1)}\ket{\psi_{0}}$, $\dots$, $M_{r_{N}}^{(N)}...M_{r_{1}}^{(1)}\ket{\psi_{0}}$,
defines a trajectory on the Bloch sphere. This trajectory is given
by the set of geodesics connecting the states. For Hermitian Kraus
operators, $M_{r_{j}}^{(j)\dagger}=M_{r_{j}}^{(j)}$ (which is the
case in Eq.~(\ref{eq:partial})), the measurement-induced phase,
(\ref{eq:phase}), has a geometric interpretation as $\chi_{\{r_{j}\}}=\Omega/2$,
where $\Omega$ is the solid angle subtended by the trajectory on
the Bloch sphere~\citep{Cho2019,Gebhart2020,Snizhko2021,Snizhko2021a}.

\subsection{\label{subsec:top-trans}Protocol for topological transition in measurement-induced
phases}

Consider a family of measurement sequences, defined in the main text.
Each sequence consists of $N$ measurements corresponding to $\left(\theta,\phi_{j}=2\pi j/(N+1)\right)$,
where $j=1,...,N$ is the measurement number. The family is obtained
when considering such sequences at all $\theta\in[0,\pi]$. We perform
postselection, which restricts measurement readouts to be $r_{j}=-$
for all $j$. According to Eq.~(\ref{eq:phase}), this defines a
phase $\chi_{\left\{ -\right\} }(\theta)$ for each measurement sequence
in the family. For briefness, we denote this phase $\chi(\theta)$.

The function $\chi(\theta)$ possesses a topological invariant. This
follows from the fact that at $\theta=0$ and $\theta=\pi$, $M_{r_{j}}^{(j)}\ket{\psi_{0}}=\ket{\psi_{0}}$
for all $j$, implying that $\ket{\psi_{f}}=\ket{\psi_{0}}$. That
is, the measurements do not change the system state, so that the resulting
phase is trivial: $e^{i\chi(0)}=e^{i\chi(\pi)}=1$. As any phase,
the measurement-induced phase is defined up to an integer multiple
of $2\pi$. Without loss of generality, we can set $\chi(0)=0$.
This, however, eliminates the freedom of adding multiples of $2\pi$
at all other $\theta$ due to the natural demand of continuity of
$\chi(\theta)$. In particular, $\chi(\pi)$ may be non-zero; yet
$e^{i\chi(\pi)}=1$ implies that $\chi(\pi)=2\pi m$ with integer
$m$. In other words, the difference $\Delta\chi=\chi(\pi)-\chi(0)$,
must be quantized in units of $2\pi$, as stated in the main text.
Further, the quantization of $\Delta\chi$ implies that its value
cannot be changed by continuous deformations of the function $\chi(\theta)$.
Therefore, $\Delta\chi$ constitutes a topological invariant.

As discussed in the main text, for infinitely weak ($\zeta\to0$)
and projective ($\zeta\to1$) measurements $\Delta\chi_{\zeta\rightarrow0}=0$
and $\Delta\chi_{\zeta\rightarrow1}=2\pi$ respectively. This necessitates
a jump in the topological invariant, i.e., a topological transition,
at some critical measurement strength, $\zeta_{c}\in[0,1]$.

\subsection{Detection of measurement-induced geometric phases}

In order to detect a measurement-induced phase, one needs to interfere
the state that underwent measurements with the initial unmeasured
state, as in Fig.~\ref{fig:Experimental}a. Here we provide a theoretical
description of this in the quantum measurement formalism.

The incoming photon in polarization state $\ket{\psi_{0}}$ becomes,
after the beam splitter, the state $\ket{\psi}=\frac{1}{\sqrt{2}}\ket{\psi_{0}}\otimes\left(\ket 0+\ket 1\right)$
where $\ket 0$ and $\ket 1$ refer to the two arms of the interferometer.
A sequence of measurements with Kraus operators $M_{r_{j}}\left(\mathbf{n}_{j}\right)$
is performed in the interferometer arm denoted as $\ket 0$. The arm
denoted as $\ket 1$ features only a phase shifter, $\delta$. Therefore,
after passing through the respective interferometer arms the photon
state is
\begin{equation}
\ket{\psi}_{int}=\frac{1}{\sqrt{2}}\left[\sum_{\{r_{j}\}}\ket{\Psi_{0}\left(\{r_{j}\}\right)}\otimes\ket 0+\ket{\Psi_{1}}\otimes\ket 1\right],
\end{equation}
where
\begin{align}
\ket{\Psi_{0}\left(\{r_{j}\}\right)} & =M_{r_{N}}^{(N)}...M_{r_{1}}^{(1)}\ket{\psi_{0}}\otimes\ket{\left\{ r_{j}\right\} },\\
\ket{\Psi_{1}} & =e^{i\delta}\ket{\psi_{0}}\otimes\ket{\left\{ r_{j}=-\right\} }.
\end{align}
Here we introduced the collective state of all the detectors, $\ket{\left\{ r_{j}\right\} }$. In the arm denoted by
$\ket 1$ the detectors do not interact with the photon and thus stay
in the no-click position.

The second beam splitter converts $\ket 0\rightarrow\frac{1}{\sqrt{2}}\left(\ket 0+\ket 1\right)$,
$\ket 1\rightarrow\frac{1}{\sqrt{2}}\left(\ket 0-\ket 1\right)$,
so that the output state is
\begin{align}
\ket{\psi}_{out}=\frac{1}{2}\biggl[\left(\ket{\Psi_{1}}+\sum_{\{r_{j}\}}\ket{\Psi_{0}\left(\{r_{j}\}\right)}\right)\otimes\ket 0
+\left(-\ket{\Psi_{1}}+\sum_{\{r_{j}\}}\ket{\Psi_{0}\left(\{r_{j}\}\right)}\right)\otimes\ket 1\biggr].
\end{align}
Therefore, the probability of the photon appearing at output ports
$\ket 0$ or $\ket 1$ is given by
\begin{align}
P_{0/1} & =\frac{1}{2}\left[1\pm\sum_{\{r_{j}\}}\mathrm{Re}\thinspace e^{-i\delta}\sp{\Psi_{1}}{\Psi_{0}\left(\{r_{j}\}\right)}\right]\nonumber \\
 & =\frac{1}{2}\left[1\pm\mathrm{Re}\thinspace e^{-i\delta}\bra{\psi_{0}}M_{-}^{(N)}...M_{-}^{(1)}\ket{\psi_{0}}\right].\label{eq:final_probability}
\end{align}
Note that only the term with no-click readouts, $\{r_{j}=-\}$, in
the measured interferometer arm contributes to the interference, which thus enables the observation of the measurement-induced phase,
cf.~Eq.~(\ref{eq:phase}) and \ref{subsec:phases}. In other words, the
interference implicitly performs the postselection to $\{r_{j}=-\}$
required by the definition of $\chi(\theta)$, cf.~\ref{subsec:top-trans}.

\section{\label{subsec:measurement_implementation}Optical implementation
of the null-weak measurement}

In our optical implementation of the measurements, the detectors are
not two-state systems with possible readouts $r=\pm$. We use the
photon spatial degree of freedom, i.e., its location in the $xy$
plane (transverse to the propagation direction). The formal description
of this is as follows. The incident photon's electric field can be described as
\begin{eqnarray}
\mathbf{E}_{0}(x,y)=\left(\begin{array}{c}
E_{0y}(x,y)\\
E_{0x}(x,y)
\end{array}\right) =\left(\begin{array}{c}
E_{0y}\\
E_{0x}
\end{array}\right)\sqrt{\frac{2}{\pi w_{0}^{2}}}e^{-(x^{2}+y^{2})/w_{0}^{2}}e^{i k z}.
\end{eqnarray}
The measurement is implemented via a beam displacer (see Figure~\ref{fig:Imper}a) that shifts the $x$-polarized component in space:
\begin{eqnarray}
\mathbf{E}(x,y)=\mathrm{BD}\:\mathbf{E}_{0}(x,y) =\sqrt{\frac{2}{\pi w_{0}^{2}}}\left(\begin{array}{c}
e^{ikn_{y}L_{y}}E_{0y}e^{-(x^{2}+y^{2})/w_{0}^{2}}\\
e^{ikn_{x}L_{x}}E_{0x}e^{-(\left[x-d_{x}\right]^{2}+y^{2})/w_{0}^{2}}
\end{array}\right)e^{i k z}.
\end{eqnarray}
Apart from the displacement, the phases associated with propagation
in the beam displacer, $kn_{x}L_{x}$ and $kn_{y}L_{y}$, are imprinted
onto the polarization components. The overall phase is not important, whereas
the difference $\gamma=kn_{x}L_{x}-kn_{y}L_{y}$ may lead to observable
consequences, cf.~Fig.~\ref{fig:phase_diagram}. In our protocol
we compensate this phase difference, see below. Therefore, here we
put, for simplicity, $kn_{x}L_{x}=kn_{y}L_{y}=0$.

If, after experiencing the BD, the beam were to interfere with the
original beam, the interference term would be
\begin{eqnarray}
\int dx\,dy\:\mathbf{E}_{0}^{*}(x,y)\mathbf{E}(x,y)
=\abs{E_{0y}}^{2}+\abs{E_{0x}}^{2}e^{-d^{2}/(2w_{0}^{2})}
=\left(\begin{array}{c}
E_{0y}\\
E_{0x}
\end{array}\right)^{\dagger}M_{-}\left(\begin{array}{c}
E_{0y}\\
E_{0x}
\end{array}\right),
\end{eqnarray}
where $M_{-}$ is the matrix defined in Eq.~(\ref{eq:partial}) if
$\ket{\uparrow}$ is interpreted as the $y$ polarization and $\ket{\downarrow}$
as the $x$ polarization, while $\sqrt{1-\zeta}=e^{-d_{x}^{2}/(2w_{0}^{2})}$.
Therefore, a beam displacer implements a postselected null weak measurement
in the photon's polarization space. The measurement strength $\eta=\sqrt{-\ln\left(1-\zeta\right)}=d_{x}/w_{0}$,
as defined in the main text. The limit of projective measurement corresponds to $\eta\rightarrow\infty$, while
the infinitely weak measurement corresponds
to $\eta\rightarrow0$.

Note that in our actual setup, cf.~Fig.~\ref{fig:Experimental}a,
the interference happens after three beam displacements have been
performed. Therefore, the postselection is implemented not on the
readout of each individual measurement, but on the combined ``readout''
of all measurements. This consitutes an important conceptual difference
compared to the original definition of the measurement-induced phase
and its topological transition, cf.~\ref{subsec:phases} and
\ref{subsec:top-trans}. Observation of the topological transition
in our work, thus, underlines that the transition is not a feature
of a specific narrow protocol, but a more general phenomenon.

\emph{Phase difference compensation.} In order to compensate the unwanted
phase difference $\gamma=kn_{x}L_{x}-kn_{y}L_{y}$, one can employ
a phase plate,
\begin{equation}
\text{P}(\varphi)=\begin{pmatrix}e^{i\varphi/2} & 0\\[1.5ex]
0 & e^{-i\varphi/2}
\end{pmatrix}.
\end{equation}
Choosing $\varphi=\gamma$ and placing the phase plate after the beam
displacer leads to
\begin{eqnarray}
\text{P}(\gamma)\mathrm{\mathrm{\:BD\:}}\mathbf{E}_{0}(x,y)=\sqrt{\frac{2}{\pi w_{0}^{2}}}e^{ik(n_{x}L_{x}+n_{y}L_{y})/2}
\times\left(\begin{array}{c}
E_{0y}e^{-(x^{2}+y^{2})/w_{0}^{2}}\\
E_{0x}e^{-(\left[x-d_{x}\right]^{2}+y^{2})/w_{0}^{2}}
\end{array}\right)e^{i k z},
\end{eqnarray}
leaving one only with an unimportant overall phase. The overall phase
is unimporant because it does not depend on the incoming polarization,
and thus can be calibrated away.

In our setup, cf.~Fig.~\ref{fig:Experimental}a, the required phase
compensation is implemented with a quarter wave plate for a wave length
distinct from that of the laser we employ. We denote it as CWP.

\emph{Measuring different observables.} The measurement procedure described
above leads to the back action matrix $M_{-}$ as defined in Eq.~(\ref{eq:partial}),
i.e., to measuring $\sigma_{z}$. In order to implement measurements
of different observables $\mathbf{n}\cdot\bm{\sigma}$, corresponding
to $\mathbf{n}=(\sin\theta\cos\phi,\sin\theta\sin\phi,\cos\theta)$,
one needs to be able to (i) discriminate different linear polarizations
(not only horizontal and veritcal) with a beam displacer and (ii)
convert elliptical polarizations to linear and back, so that they
can be discriminated by the beam displacer.

(i) can be implemented by rotating the beam displacer in the $xy$
plane:
\begin{equation}
\mathrm{BD}(\theta/2)=R(\theta/2)\mathrm{\:BD\:}R(-\theta/2),
\end{equation}
with the rotation matrix 
\begin{equation}
R(\theta/2)=\begin{pmatrix}\cos\theta/2 & -\sin\theta/2\\
\sin\theta/2 & \cos\theta/2
\end{pmatrix}.
\end{equation}

(ii) can be implemented by placing phase plates $P(\pm\phi)$ before
and after the beam displacer.

Therefore, a measurement of $\mathbf{n}\cdot\bm{\sigma}$ as described
in \ref{subsec:measurement} can be implemented via a sequence
of elements that involves a rotated BD and CWP, as well as two phase
plates:
\begin{equation}
\mathcal{M}(\theta,\phi)=\mathrm{P}(-\phi)R(\theta/2)\text{P}(\gamma)\mathrm{\:BD\:}R(-\theta/2)\mathrm{P}(\phi).\label{eq:measurement_stage_naive}
\end{equation}
Note that in order to rotate the measurement axis by $\theta$, one
needs to perform real space rotations by $\alpha=\theta/2$.

This sequence uses four elements per measurement, whereas our setup
in Fig.~\ref{fig:Experimental} features only three optical elements
per measurement. We describe how this is achieved in the next section.\\

\subsection{Simplifying the experimental setup}
\label{SI:A6}
The protocol for observing the topological transition requires sending
in a laser beam with polarization
\begin{equation}
\mathbf{E}_{in}=\left(\begin{array}{c}
E_{0y}\\
E_{0x}
\end{array}\right)=\left(\begin{array}{c}
\cos\theta/2\\
\sin\theta/2
\end{array}\right)=R(\theta/2)\left(\begin{array}{c}
1\\
0
\end{array}\right).
\end{equation}
and using $N$ measurements $\mathcal{M}(\theta,\phi_{j})$, where
the measurement stages are defined in Eq.~(\ref{eq:measurement_stage_naive})
and $\phi_{j}=2\pi j/(N+1)$. The number of required optical elements
can be reduced. In order to do this, one needs two observations.

First, consider the incoming polarization and the first measurement:
\begin{eqnarray}
\mathcal{M}(\theta,\phi_{1})\left(\begin{array}{c}
\cos\theta/2\\
\sin\theta/2
\end{array}\right)
=\mathrm{P}(-\phi_{1})R(\theta/2)\mathrm{\text{P}(\gamma)}\mathrm{\:BD\:}\underbrace{R(-\theta/2)\mathrm{P}(\phi_{1})R(\theta/2)}\left(\begin{array}{c}
1\\
0
\end{array}\right).
\end{eqnarray}
The block $R(-\theta/2)\mathrm{P}(\phi_{1})R(\theta/2)$ can be interpreted as a phase plate rotated by the angle $\alpha=\theta/2$, $\mathrm{P(\phi_{1},\alpha)}$.

Second, consider two sequential measurements
\begin{equation}
\mathcal{M}(\theta,\phi_{j+1})\mathcal{M}(\theta,\phi_{j})=\mathrm{P}(-\phi_{j+1})R(\theta/2)\mathrm{\text{P}(\gamma)}\mathrm{\:BD\:}\underbrace{R(-\theta/2)\mathrm{P}(\phi_{j+1})\mathrm{P}(-\phi_{j})R(\theta/2)}\mathrm{\text{P}(\gamma)}\mathrm{\:BD\:}R(-\theta/2)\mathrm{P}(\phi_{j}).
\end{equation}

The block $R(-\theta/2)\mathrm{P}(\phi_{j+1})\mathrm{P}(-\phi_{j})R(\theta/2)$
can be replaced with a single rotated phase plate $\mathrm{P}(\phi_{j+1}-\phi_{j},\alpha)=\mathrm{P}(2\pi/(N+1),\alpha)=\mathrm{P}(\phi_{1},\alpha)$.

Therefore, instead of having a rotated incoming polarization and rotated
beam displacers, one can have vertical incoming polarization and rotated
phase plated $\mathrm{P}(\phi_{1},\alpha)$ before the beam displacers.
Note that this setup simplification involves replacing all phase plates
$\mathrm{P}(\phi_{j})$ with their rotated versions $R(-\theta/2)\mathrm{P}(\phi_{j})R(\theta/2)$
and the input polarization $R(\theta/2)\begin{pmatrix}1 & 0\end{pmatrix}^{\mathrm{T}}$
with $\begin{pmatrix}1 & 0\end{pmatrix}^{\mathrm{T}}$. The simplified setup is related to the orignial protocol by rotating
all the measurement axes $\mathbf{n}_{j}$ by angle $\theta$ around
the $y$ axis of the Bloch sphere.

For our choice of $N=3$, we have $\phi_{1}=\pi/2$, making the required phase
plates $\mathrm{P}(\phi_{1},\alpha)$ quarter wave plates and leading
to the setup in Fig.~\ref{fig:Experimental}a.

\subsection{Effect of imperfect birefringent crystals \label{SI:Imper}}

\begin{figure}
\begin{centering}
\includegraphics[width=0.8\columnwidth]{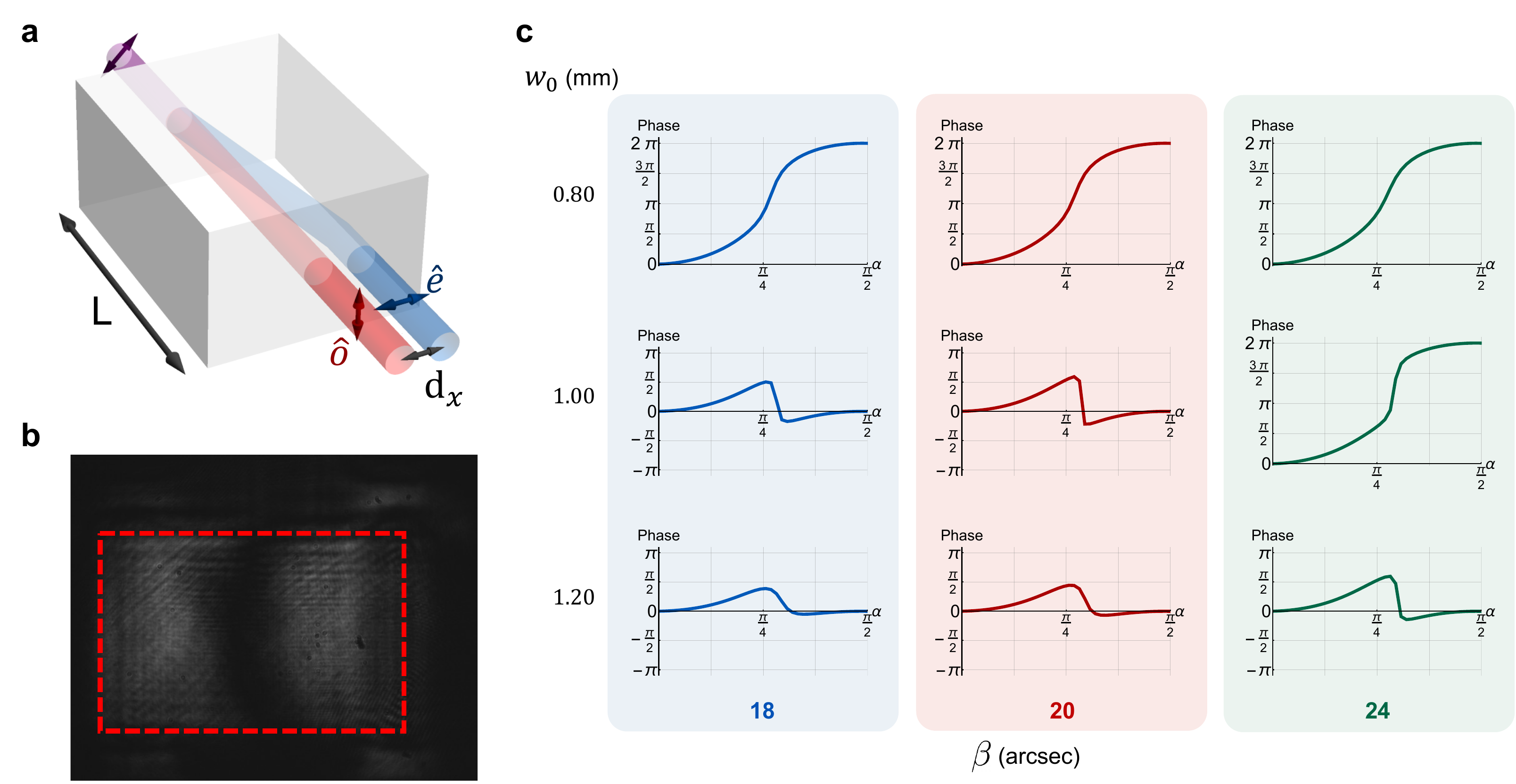}
\par\end{centering}
\caption{\label{fig:Imper} \textbf{Modeling an imperfect beam displacer. a. }Schematic of a beam displacer (BD). \textbf{b.} Interference pattern obtained from the output beams after projecting on circular polarization (and with a circularly polarized input beam). \textbf{c. } Geometrical phase curves for different values of the deviation angle  $\beta$.  }
\end{figure}

In our previous description, we have assumed that both the extraordinary and ordinary components of the field exit from the birefringent crystal along parallel directions. Nevertheless, while extremely small, the beam displacer (THORLABS BDY12) manufacturer mentions that both components are parallel to each other within 30 arcseconds. Figure \ref{fig:Imper}-b shows the intensity profile from a YBO$_4$ crystal when illuminated with diagonally polarized light, and a polarizer is placed on the output. The presence of an interference pattern corroborates the existence of a small deviation in the propagation direction of one of the output beams. 

Following these results, we modified the model for our beam displacer to account for our crystals' imperfections. First, we assumed that only the extraordinary component experiences a deflection from the optical axis by an angle $\beta$. Due to variability from one crystal to another, we allow the deviation to occur along any direction in the transverse plane, characterized by an angle $\nu$. Therefore, it is possible to write the effect of an imperfect beam displacer on an impinging electric field as 

\begin{eqnarray}
\mathbf{E}(x,y)=\mathrm{BD}_{\nu,\beta}\:\mathbf{E}_{0}(x,y) =\sqrt{\frac{2}{\pi w_{0}^{2}}}\left(\begin{array}{c}
e^{ikn_{y}L_{y}}E_{0y}e^{-(x^{2}+y^{2})/w_{0}^{2}}\\
e^{ikn_{x}\Delta L_{x}}E_{0x}e^{-(\left[x-d_{x}\right]^{2}+y^{2})/w_{0}^{2}}
\end{array}\right)e^{i k z}.
\label{eq:ImpBd}
\end{eqnarray}

where $\Delta=\sin \beta (\cos\nu x+\sin\nu y)$ implements the deflection on the extraordinary component. Therefore, it is possible to describe the evolution of the input beam $\mathbf{E}_0(x,y)$ due to a sequence of $N$ stages with distinct imperfections as 

\begin{equation}
    \mathbf{E_u}(x,y)=\prod_{j=1}^N\mathcal{M}_{\nu_j,\beta_j}(\theta,\phi_j)\mathbf{E}_0(x,y),\label{eq:NewStage}
\end{equation}
where $\mathcal{M}_{\nu,\beta}(\theta,\phi)$ is obtaining by substituting Eq.~\eqref{eq:ImpBd} into Eq.~\ref{eq:measurement_stage_naive}. At first glance at the effect of these imperfections on the location of the transition, let us consider the case when the $N=3$ measurement stages possess identical crystals. As shown in Figure~\ref{fig:Imper}c, the location of the topological transition shifts towards higher values of $w_0$ when the deflection angle increases.

Nevertheless, this assumption is not valid in our experiment, since each individual crystal is different. 
As a consequence, the behavior of the geometrical phase curve depends directly on three pairs $(\nu,\beta)$ which quantify the nonparallelism between the faces of each BD. A genetic algorithm (GA)\cite{coley1999GE} was implemented to perform the search for the set of optimal parameters $P$ that match the experimental results.

\end{document}